\documentclass[aps,prl,twocolumn,10pt,a4paper,superscriptaddress,notitlepage, longbibliography]{revtex4-2}

\usepackage{amsmath}
\usepackage{amssymb}
\usepackage{textcomp}
\usepackage{units}
\usepackage{graphicx}
\usepackage{microtype}
\usepackage{enumitem}
\usepackage{xcolor}
\usepackage{multirow}   
\usepackage{physics}
\usepackage{siunitx}
\usepackage{booktabs}
\usepackage{csquotes}

\usepackage{hyperref}
\hypersetup{colorlinks=true, citecolor=blue, urlcolor=blue, linkcolor=blue}

\usepackage{txfonts}

\newcommand{\vect}[1]{\ensuremath{\boldsymbol{#1}}}
\newcommand{\matr}[1]{\ensuremath{\underline{#1}}}
\newcommand{\trpvect}[1]{\ensuremath{\boldsymbol{#1}^\intercal}}

\newcommand{\mub}{\mu_\mathrm{B}}
\newcommand{\kb}{k_\text{B}}

\newcommand{\mathpi}{\piup}

\newcommand{\hamil}{\mathcal{H}}
\newcommand{\hmatr}{\matr{\mathcal{H}}}
\newcommand{\adj}[1]{#1^\dagger}
\newcommand{\metric}{\matr{G}}

\usepackage{pifont}

\begin{document}
\title{Thermal Hall effect of magnons in collinear antiferromagnetic insulators:\\ signatures of magnetic and topological phase transitions}

\author{Robin R.~Neumann}
\affiliation{Institut f\"ur Physik, Martin-Luther-Universit\"at Halle-Wittenberg, D-06099 Halle (Saale), Germany}
\author{Alexander Mook}
\affiliation{Department of Physics, University of Basel, Klingelbergstrasse 82, CH-4056 Basel, Switzerland}
\author{J\"urgen Henk}
\affiliation{Institut f\"ur Physik, Martin-Luther-Universit\"at Halle-Wittenberg, D-06099 Halle (Saale), Germany}
\author{Ingrid Mertig}
\affiliation{Institut f\"ur Physik, Martin-Luther-Universit\"at Halle-Wittenberg, D-06099 Halle (Saale), Germany}

\begin{abstract}
	We demonstrate theoretically that the thermal Hall effect of magnons in collinear antiferromagnetic insulators is an indicator of magnetic and topological phase transitions in the magnon spectrum. The transversal heat current of magnons caused by a thermal gradient is calculated for an antiferromagnet on a honeycomb lattice. An applied magnetic field drives the system from the antiferromagnetic phase via a spin-flop phase into the field-polarized phase. Besides these magnetic phase transitions we find topological phase transitions within the spin-flop phase. Both types of transitions manifest themselves in prominent and distinguishing features in the thermal conductivities; depending on the temperature, the conductivity changes by several orders of magnitude, providing a tool to discern experimentally the two types of phase transitions. We include numerical results for the van der Waals magnet MnPS$_3$.
\end{abstract}
\date{\today}
\maketitle

\paragraph{Introduction.}
In electronic systems, details of the electronic structure and of the magnetic configuration manifest themselves in the transport properties.
As an example, the quantum anomalous Hall effect, in which the transversal transport coefficient is quantized, is a clear signature of a topologically nontrival phase.
Moreover, topological phases of the electronic states can be clearly identified spectroscopically, e.\,g. in topological insulators \cite{Hsieh2008,Mourik2012,Qi2011,Hasan2010,Li2016Experimental}.

The field of topology is not restricted to fermions, but also applies to bosons. The topological features of phonons \cite{Prodan2009,Zhang2010,Berg2011,Qin2012,Chen2016,Xia2017,Akazawa2020}, photons \cite{Chiao1986,Lu2014,Lu2016,Ozawa2019,Chen2019}, and magnons \cite{Zhang2013,Mook14a,Mook14b,owerre2017floquet,McClarty2018,Rueckriegel2018,Diaz2019,Wang2020}, however, are more subtle due to the lack of the Pauli exclusion principle which results in quantized transport. In this Paper we focus on magnons, because they are easily manipulated by external magnetic fields. The identification of magnon edge states, the hallmarks of a nontrivial system, is notoriously difficult. On the one hand, angle-resolved photo\-electron and spin-polarized scanning tunneling spectroscopy cannot be applied at all or not without severe restrictions \cite{Pietzsch2001,Bode2007,FernandezRossier2009,Fransson2010,Balashov2006,Feldmeier2020}. On the other hand, inelastic neutron scattering succeeds in detecting gapped bulk spectra, but fails in resolving edge modes \cite{Zhu2021}. These apparent shortcomings call for identifying clear signatures of magnetic and topological phase transitions in the magnetotransport of magnons, for example in the thermal Hall conductivity.

In this Paper, we aim at bridging the apparent gap sketched in the preceding paragraph. For this purpose we investigate theoretically an antiferromagnet that exhibits spin-split, nonreciprocal magnon bands and both magnetic and topological phase transitions induced by an applied magnetic field. These phase transitions show up as clear characteristic signatures in the field and temperature dependence of the thermal Hall conductivity, that are explained by the magnonic band structure and the Berry curvature.
We exemplarily calculate the thermal Hall magnetoconductivity at two phase transitions, which quantifies the influence of the transitions on the Hall response, to convey the strong tunability and sensitivity of the thermal Hall effect.
Our findings suggest a means for identifying magnetic and topological phases via transport measurements.
Conversely, they insinuate a way to externally control the thermal Hall effect due to the significant changes across the phase transitions.
The numerical results for MnPS$_3$, which is known for its nontrivial magnon transport \cite{Shiomi2017}, ask for comparison with experimental data.

Previous reports addressed thermal Hall effects in collinear ferromagnets with Dzyaloshinskii-Moriya interaction (DMI) and dipolar interactions \cite{Katsura2010,Onose2010,Matsumoto2011,Matsumoto2011a,Shindou13,Zhang2013,Ruckriegel2018,Ideue2012,Hirschberger2015,Mook14a,Matsumoto2014,Cao2015,Lee2015,Xu2016,Nakata2017,Owerre2016,Owerre2016a,Owerre2016b,Owerre2016c,Mook2016c,Madon2016,Okamoto2017,Murakami2017,Owerre2017,Li2018,Seshadri2018,kawano2018thermal}, in weak ferromagnets with scalar spin chirality or due to magnetic fields \cite{Hirschberger2015b,Hoogdalem2013,Schuette2014,Mook2017a,Owerre2017a,Owerre2017b,Owerre2017c,owerre2017noncollinear,Iacocca2017,hwang2017magnon,Laurell2017,Laurell2018,Owerre2018,Cookmeyer2018,kim2018tunable}, in noncollinear antiferromagnets \cite{Mook2019} or in paramagnets \cite{Lee2015,Hirschberger2015b,Hirschberger2015,Watanabe2016,Kasahara2018,Kasahara2018b,Doki2018,Hentrich2019,Yamashita2019,Akazawa2020}. Here, we present a thermal Hall effect in collinear antiferromagnets without DMI, which may even be present without external fields.
While noncollinear antiferromagnets rely \emph{exclusively} on their magnetic order to break an effective time-reversal symmetry (which is a prerequisite for the thermal Hall effect), collinear antiferromagnets need nonmagnetic atoms \emph{in addition} to break it.
The underlying mechanism is the magnonic analogue of the Hall effect reported in Ref.~\onlinecite{Smejkal2020}.

\begin{figure}
    \centering
    \includegraphics[width=0.8\columnwidth]{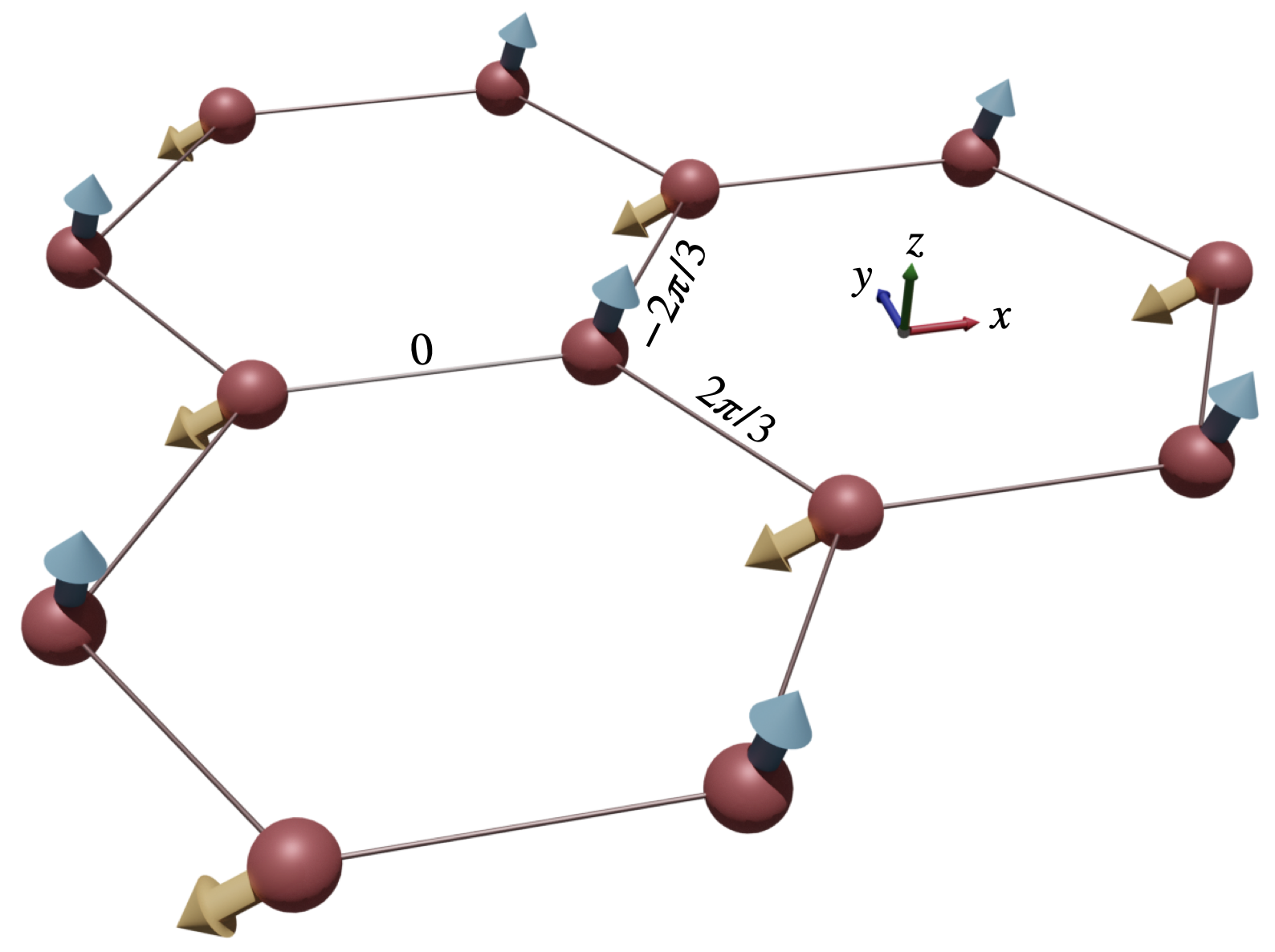}
    \caption{Honeycomb lattice with antiferromagnetically coupled spins on sublattices A (blue) and B (orange). The spin configuration shown here is paradigmatic for the spin-flop phase with $A \neq 0$ and a magnetic field applied along $-z$.
    }
    \label{fig:lattice}
\end{figure}

\paragraph{Model and methods.}
We consider a magnet on a two-dimensional (2D) honeycomb lattice (in the $xy$ plane; depicted in Fig.~\ref{fig:lattice}). In the ground state without a magnetic field the spins of sublattice~A (B) point in $+z$ ($-z$) direction.

The spin Hamiltonian
\begin{align}
	\hamil &= \hamil_\text{nn} + \hamil_\text{on} + \hamil_{B}
\label{eq:hamilExc}
\end{align}
comprises the coupling of nearest-neighbor spins,
\begin{align}
	\hamil_\text{nn} &= \frac{1}{2\hbar^2} \sum_{\langle ij \rangle}
	\trpvect{S}_i
	\begin{pmatrix}
		J + J_a \cos\theta_{ij} & -J_a \sin\theta_{ij} & 0 \\
		-J_a \sin\theta_{ij} & J - J_a \cos\theta_{ij} & 0 \\
		0 & 0 & J_z
	\end{pmatrix}
	\vect{S}_j.
\label{eq:hamilNN}
\end{align}
Both in- and out-of-plane spin components are coupled antiferromagnetically, but with different strengths ($J_z > J > 0$). The traceless and symmetric coupling, introduced by $J_a$, originates from spin-orbit coupling (SOC) \cite{Matsumoto2020}. It is related to the nearest neighbor bonds $\langle ij \rangle$ by the bond-dependent angles $\theta_{ij} = 0$, $2\mathpi/3$, and $-2\mathpi/3$ (cf. angles near bonds in Fig.~\ref{fig:lattice}). The classical collinear configuration favored by $J$ and $J_z$ is maintained as long as $J_a$ is sufficiently small.

This model was proposed for manganese thiophosphate MnPS$_3$ in Ref.~\onlinecite{Matsumoto2020} and produces a nonreciprocal magnon spectrum. The nonreciprocity may also be caused by DMI (which would not cause the spin splitting of the bands) \cite{Cheng2016} and dipole-dipole interactions \cite{Pich1995}. An experimental work by Wildes \textit{et al.} did not reveal convincing signatures of an asymmetric band structure, however, (i) based on the band splitting DMI was ruled out in MnPS$_3$ \cite{Wildes2021} (ii) bond-dependent exchange interaction $J_a$ is allowed by symmetry \cite{Matsumoto2020}, and (iii) while $J_a$ causes nonreciprocal magnon bands, it cannot be excluded due to the limited experimental resolution. In favor of neglecting DMI we recall that the bond-dependent exchange reproduces the band splitting reasonably well. Nevertheless, further insights into the spin-spin interactions are desirable, for example by comparing experimental results with the predictions  for the transport properties reported here.

We extend the model of Ref.~\onlinecite{Matsumoto2020} by considering an on-site anisotropy
\begin{align}
	\hamil_\text{on} &= -\frac{A}{\hbar^2} \sum_{i \in \text{A}} \qty(S_i^z)^2
\label{eq:hamilON}
\end{align}
for the spins on sublattice~A, which breaks the inversion symmetry on the level of the Hamiltonian. It may be brought about by placing the sample on a substrate or in a heterostructure [e.\,g., on a transition-metal dichalcogenide (TMDC)], thereby producing local environments of the atoms that differ for the two sublattices \cite{HidalgoSacoto2020}. The anisotropy translates into a sublattice-dependent on-site potential of the magnons.

The Zeeman Hamiltonian 
\begin{align}
	\hamil_{B} &= \frac{g \mub B_z}{\hbar} \sum_{i} S_i^z
\label{eq:hamilB}
\end{align}
accounts for an out-of-plane magnetic field that destabilizes the antiferromagnetic (AFM) order and induces magnetic phase transitions. 
Below the critical magnetic field $B^\text{(m)}_1$, with energy
\begin{align}
	g \mub B_1^\text{(m)} / S &= \sqrt{(3 J_z + K)^2 - 9 J^2} - K,
\end{align}
the classical ground state is a collinear antiferromagnet with a N\'eel vector pointing in $z$ direction.
Between $B_1^\text{(m)}$ and $B_2^\text{(m)}$, 
\begin{align}
	g \mub B_2^\text{(m)} / S &= 3 J_z + \sqrt{9 J^2 + K^2} - K.
\end{align}
the system is in a coplanar spin flop (SF) phase, and in the field-polarized (FP) phase (fields larger than $B_2^\text{(m)}$) all spins point along $+z$. The ground state's spin configuration has been obtained by analytical and numerical methods; for details see the Supplementary Material (SM)~\cite{Supplement}.

A Holstein-Primakoff transformation rewrites the spin operators $\vect{S}_i$ in terms of bosonic operators $a_{i}$ and $a^\dagger_{i}$ \cite{Holstein1940}. After a Fourier transformation the Hamiltonian with magnon-magnon interactions neglected reads 
$
	\hamil_2 = \frac{1}{2} \sum_{\vect{k}} \adj{\vect{\psi}}_{\vect{k}} \hmatr_{\vect{k}} \vect{\psi}_{\vect{k}},
$
with $\adj{\vect{\psi}}_{\vect{k}} = \mqty(\adj{a}_{1, \vect{k}} & \adj{a}_{2, \vect{k}} & a_{1, -\vect{k}} & a_{2, -\vect{k}})$ and wave vector  $\vect{k}$. In order to diagonalize the matrix $\hmatr_{\vect{k}}$ we introduce the matrix $\matr{T}_{\vect{k}}$ which satisfies
$
	\matr{\mathcal{E}}_{\vect{k}} = \adj{\matr{T}}_{\vect{k}} \hmatr_{\vect{k}} \matr{T}_{\vect{k}}
	,\ 
	\metric = \adj{\matr{T}}_{\vect{k}} \metric \matr{T}_{\vect{k}}.
$
Here, the diagonal matrix $\matr{\mathcal{E}}_{\vect{k}}
 = \mathrm{diag}(\mqty{
 \varepsilon_{1,\vect{k}} & \varepsilon_{2,\vect{k}} &
 \varepsilon_{1,-\vect{k}} & \varepsilon_{2,-\vect{k}}}
 )$ contains the magnon energies $\varepsilon_{n,\vect{k}}$ of both bands ($n = 1, 2$).  $\metric = \mathrm{diag}(\mqty{1 & 1 &-1 &-1})$ is the bosonic metric.

The topological phases of the magnon spectrum are characterized by the Chern numbers
$
	C_n = -\frac{1}{2 \mathpi}
	\int_{\text{1. BZ}} \varOmega_{n\vect{k}} \,\mathrm{d}^2 k,
$
which are integrals of the Berry curvature \cite{Mook2019}
\begin{align}
	\varOmega_{n,\vect{k}} = -2 \Im \sum_{\underset{m \neq n}{m=1}}^{4}
	\frac{\qty(\metric \adj{\matr{T}}_{\vect{k}} \partial_x \hmatr_{\vect{k}} \matr{T}_{\vect{k}})_{nm}
			  \qty(\metric \adj{\matr{T}}_{\vect{k}} \partial_y \hmatr_{\vect{k}} \matr{T}_{\vect{k}})_{mn}}{
		\qty[\qty(\metric \matr{\mathcal{E}}_{\vect{k}})_{nn} - \qty(\metric \matr{\mathcal{E}}_{\vect{k}})_{mm}]^2
	}
\end{align}
over the first Brillouin zone (BZ). Since the Chern numbers of all bands add up to zero ($C_{1} + C_{2} = 0$), it suffices to specify one Chern number, say, $C_{1}$ of the lowest band $n = 1$.

The Berry curvature also enters the thermal Hall conductivity \cite{Matsumoto2011}
\begin{align}
	\kappa_{xy} = -\frac{\kb^2 T}{\hbar V} \sum_{\vect{k}} \sum_{n=1}^N
	c_2[\rho(\varepsilon_{n,\vect{k}})] \, \varOmega_{n,\vect{k}},
	\label{eq:thermalHall}
\end{align}
in which $\kb$, $T$, $V$, $\rho(x) = \qty[\exp\qty(\nicefrac{x}{\kb T}) - 1]^{-1}$ are Boltzmann's constant, the temperature, the system's volume (or area), and the Bose distribution, respectively. The magnon energies $\varepsilon_{n,\vect{k}}$ appear in
$
	c_2(x) = (1 + x) \ln^2 \frac{1 + x}{x} - \ln^2 x - 2 \mathrm{Li}_2(-x)
$
with the Spence function $\mathrm{Li}_2(z) = -\int_0^z \ln(1 - t)/t \,\mathrm{d}t$.

We continue with parameters for MnPS$_3$: next-neighbor distance $a = \SI{3.503}{\angstrom}$ \cite{Jain2013,Persson2016}, coupling strengths $J = \SI{1.54}{\milli\electronvolt}$, $J_z = \SI{1.541}{\milli\electronvolt}$, $J_a = \SI{0.02}{\milli\electronvolt}$, and spin $S = \nicefrac{5}{2}$ \cite{Wildes1998,Wildes2021}. $J_z$ is estimated by identifying the anisotropy field $g \mub H_\text{A} = \SI{8.6}{\micro\electronvolt}$ given in Ref.~\onlinecite{Wildes1998} with the strength of the (two-ion) anisotropy that is associated with $3S(J_z - J)$ (the factor of $3$ accounts for the coordination number). We neglect interlayer interactions and divide the 2D thermal conductivity by the interlayer distance of \SI{7.278}{\angstrom} to obtain the conductivities of the three-dimensional (3D) system \cite{Jain2013,Persson2016}.

Regarding the on-site anisotropy $A$, we consider two cases. First, the bulk properties of MnPS$_3$ are modeled by setting $A = 0$. Second, we account for a substrate by setting $A = \SI{0.1}{\milli\electronvolt}$, which is a realistic value in the range of predictions by ab initio calculations for other van der Waals magnets \cite{HidalgoSacoto2020}. Our choice for $A$ renders the respective calculations semiquantitative, since the precise numerical value of $A$ depends presumably on the selected substrate.

\begin{figure}
	\includegraphics[width=\linewidth]{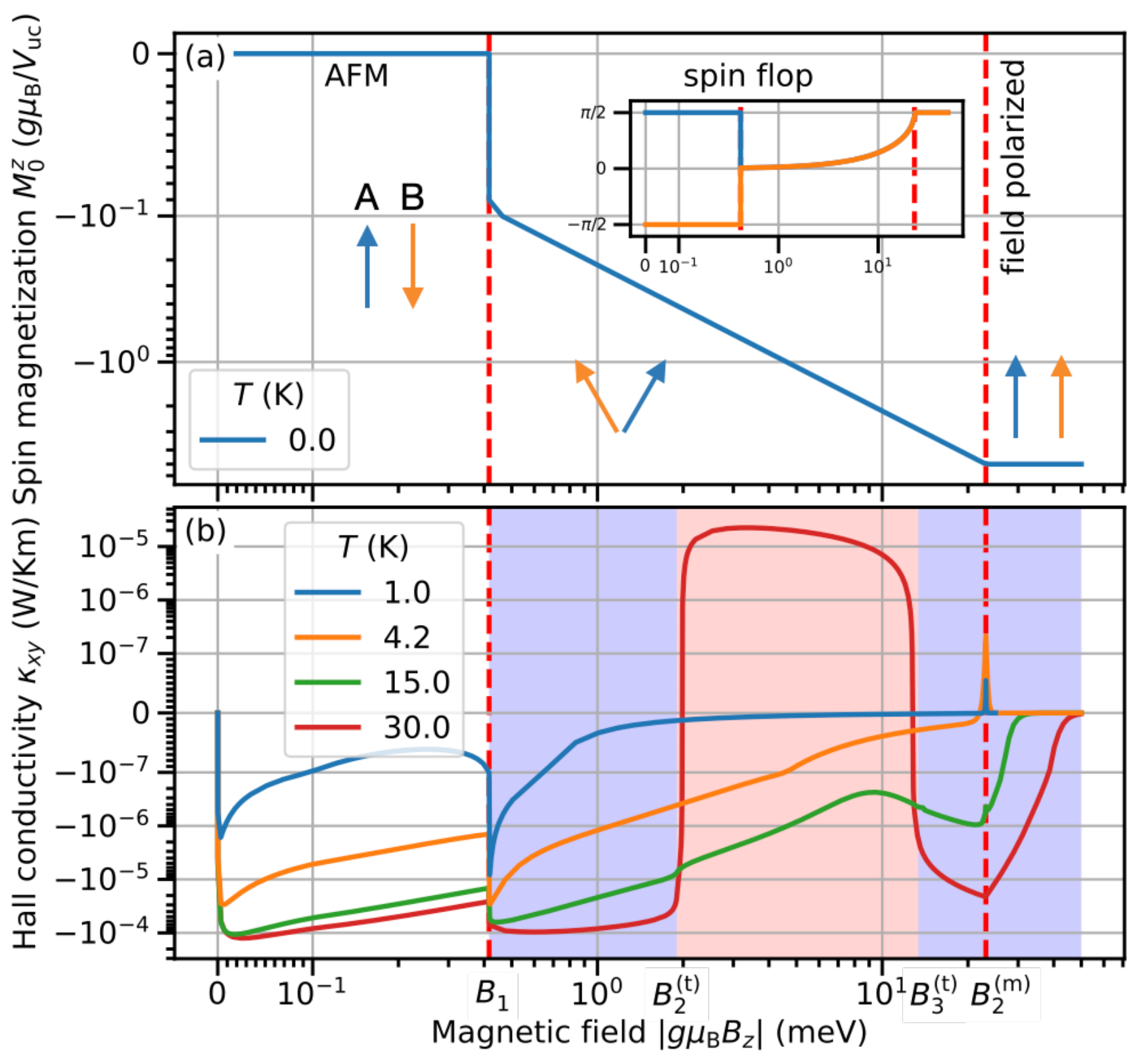}
	\caption{Magnetic, topological, and transport properties of (bulk) MnPS$_{3}$ ($A = 0$). (a) Classical ground state magnetization versus the magnetic field.
	Inset: angles $\theta_{\text{A}}$ and $\theta_{\text{B}}$ of the sublattice A (blue) and B (orange) spins with the $xy$ plane.
	(b) Thermal Hall conductivity $\kappa_{xy}$ for four selected temperatures ($T =$ \SI{1.0}{\kelvin},  \SI{4.2}{\kelvin}, \SI{15}{\kelvin}, and \SI{30}{\kelvin}). The white/blue/red background color indicates  topological phases with Chern numbers $C_{1} = 0, -1, +1$ of the lowest magnon band.
	Dashed red lines mark the magnetic phase transitions at the critical fields $B^\text{(m)}_{1}$ and $B^\text{(m)}_{2}$.
	All four panels have logarithmic ordinates and abscissae with linear-scale segments around $0$, which are identified by equally spaced minor ticks.
	For parameters see text.
 \label{fig:the}}
\end{figure}

\begin{figure}[h]
	\includegraphics[width=.99\linewidth]{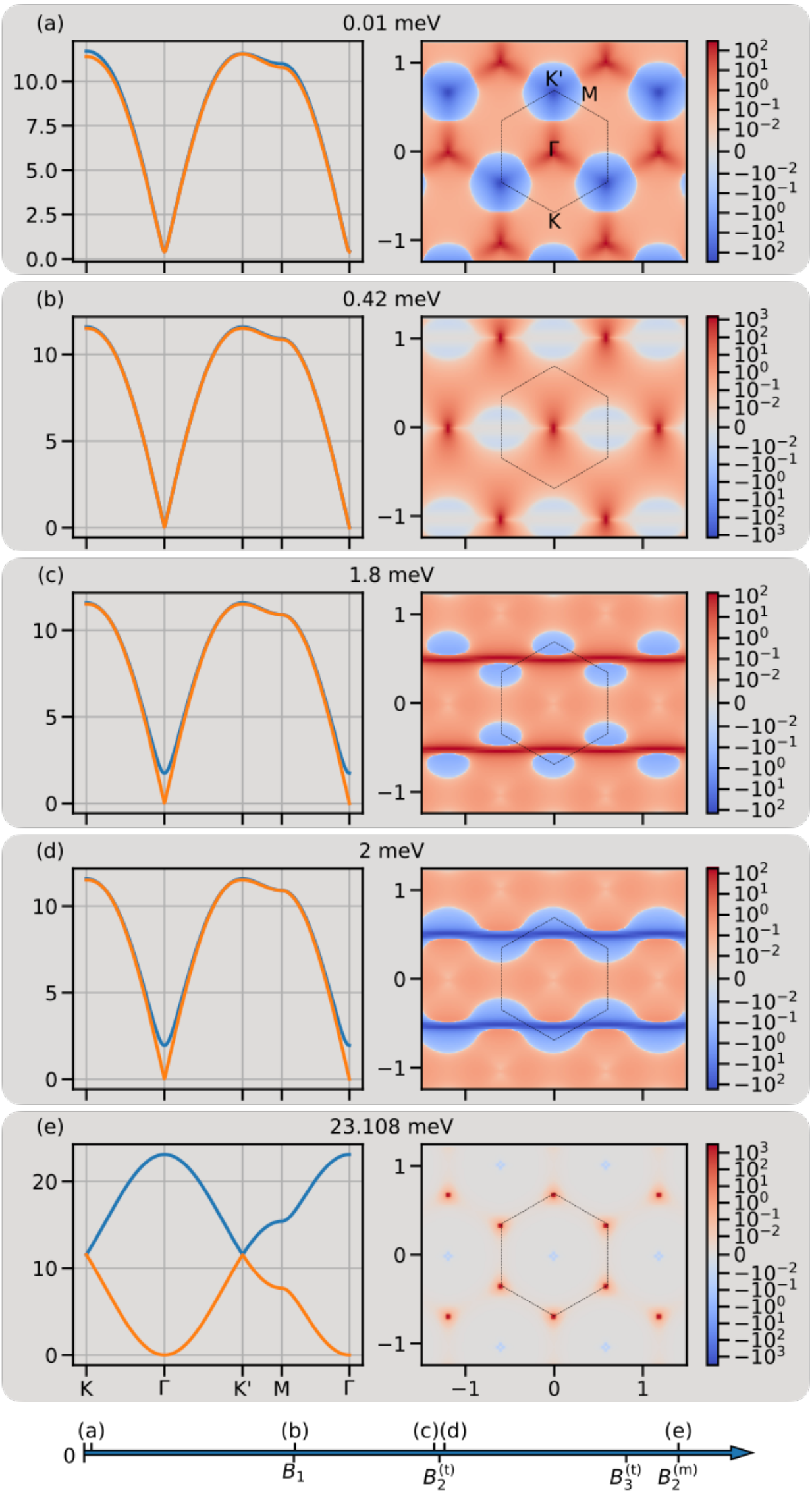}
	\caption{Magnon band structures and Berry curvatures of (bulk) MnPS$_{3}$ ($A = 0$) for selected strengths $|g \mub B_z|$ of the magnetic field. 
	Magnon-dispersion and Berry-curvature panels appear in pairs, indicated by a common gray background, with identical strength of the magnetic field (in $\unit{meV}$; the positioning with respect to the phase transitions is sketched at the bottom). The magnon energies $\varepsilon_{n\vect{k}}$ (in \si{\milli\electronvolt}) are shown along high-symmetry lines of the first Brillouin zone; the Berry curvatures $\varOmega_{1\vect{k}}$ of the lowest band are displayed as color maps in reciprocal space (the black hexagons indicate the first BZ\@). The $k_x$ and $k_y$ axes are given in \si{\per\angstrom}. Parameters are chosen as in Fig.~\ref{fig:the}. \label{fig:spectra_berryc}}
\end{figure}

\begin{table}
    \centering
    \caption{Critical fields at which topological (t) and magnetic (m) phase transitions occur depending on the sublattice-specific anisotropy~$A$. All values in \si{\milli\electronvolt}.}
    \label{tab:critfields}
    \begin{tabular}{S[table-format=1.1] *{5}{S[table-format=2.3]}}
        \toprule \toprule
         {$A$} & {$g \mub B_1^\text{(m)}$} & {$g \mub B_2^\text{(m)}$} & {$g \mub B_1^\text{(t)}$} & {$g \mub B_2^\text{(t)}$} & {$g \mub B_3^\text{(t)}$}\\
         \cmidrule(lr{0.5em}){2-3} \cmidrule(lr{0.5em}){4-6}
         0 & 0.416 & 23.107 & 0.416 & 1.901 & 13.368 \\
         0.1 & 2.202 & 22.860 & {---} & {---} & {---} \\
         \bottomrule\bottomrule
    \end{tabular}
\end{table}

Below, we describe and explain the field-dependent Hall conductivity $\kappa_{xy}(B_z)$ for increasing field starting at zero. Magnetic (m) and topological (t) phase transitions occur at $B_1 < B^\text{(t)}_2 < B^\text{(t)}_3 < B^\text{(m)}_2$ (Table~\ref{tab:critfields}). If a topological and a magnetic phase transitions coincide (e.\,g. at $B_1$), the notation $B^\text{(m)}$ and $B^\text{(t)}$ becomes redundant. Changes in $\kappa_{xy}$ are traced back to the evolution of the magnon spectrum and the Berry curvature.

\paragraph{Discussion of results for bulk MnPS$_3$.}
For $A = 0$ and zero magnetic field, the AFM phase, in which A (B) spins point along $+z$ ($-z$), is invariant under simultaneous space inversion $P$ and time reversal $T$, which causes $\kappa_{xy} = 0$. The otherwise degenerate magnon bands are spin-orbit split by $J_a \neq 0$, with the exception of the $\Gamma$ and K' points in the BZ \cite{Matsumoto2020}. The Berry curvature vanishes because of the aforementioned symmetry.

A small magnetic field breaks this symmetry and lifts the band degeneracies at $\Gamma$ and K'. The resulting increase of $\kappa_{xy}$ (in absolute value) with temperature [cf.~Fig.~\ref{fig:the}(b)] is explained by the magnon band structure and the Berry curvature. As an example we discuss the case $|g \mub B_z| = \unit[0.01]{meV}$ [Fig.~\ref{fig:spectra_berryc}(a)]. The lifting of the band degeneracies at $\Gamma$ [$\varOmega_{n\vect{k}} > 0$, red in the right panel of (a)] and K' ($\varOmega_{n\vect{k}} < 0$) brings about Berry curvature of opposite sign. The higher thermal occupation of the states around $\Gamma$ and the minus sign in Eq.~\eqref{eq:thermalHall} explain that $\kappa_{xy}$ is negative. The higher the temperature, the larger the occupation at $\Gamma$ and the larger $|\kappa_{xy}|$. The three-fold rotational symmetry of the spin-lattice is reflected in the Berry curvature.

Magnons belonging to the lower (upper) band are located mostly on the B (A) sublattice, whose spins are destabilized (stabilized) by the magnetic field (applied along $-z$). Hence, the lower (upper) band is shifted to lower (higher) energies by the magnetic field. As the field strength increases, the positive Berry curvature around $\Gamma$ is gradually redistributed towards the K points and the negative Berry curvature at K' extends towards $\Gamma$ [not shown but similar to Fig.~S6(a) in SM \cite{Supplement}], which explains the non-monotonic behavior of $\kappa_{xy}$.

At the first-order AFM-SF phase transition at $g \mub B_1^\mathrm{(m)} = \SI{0.416}{\milli\electronvolt}$, also identified by a diverging susceptibility, both A and B spins are abruptly rotated into the $xy$ plane but obtain a small (ferromagnetic) component parallel to the magnetic field. This redirection is seen in the angles $\theta_\text{A}$ and $\theta_\text{B}$ between the $xy$ plane and the spins (inset: A blue, B orange) and in the jump of the magnetization from zero to negative values [Fig.~\ref{fig:the}(a)].
The experimentally measured critical field in the range of $g \mub B_1 =\text{\SIrange{0.42}{0.54}{\milli\electronvolt}}$ \cite{Okuda1986,Goossens1998} agrees reasonably well with our analysis.
Moreover, the magnetization $M_z(B_z)$ as a function of the external field reported in Ref.~\onlinecite{Okuda1986} features the same linear dependence as in Fig.~\ref{fig:the}(a) (the linear dependence remains with linear $x$ and $y$ axes) and the magnetic moment of $0.6\, \mub$ just above the transition point is close to our calculations.

In the SF phase, the lower band is pinned at zero energy at $\Gamma$ \footnote{We neglect a pseudo-Goldstone gap \cite{Rau2018} that we show to be irrelevant in the SM \cite{Supplement}.}. The symmetry of the spectrum and the Berry curvature are reduced by the spontaneous breaking of the three-fold rotational symmetry. (The Berry curvature is symmetric to the $k_x = 0$ line because of the choice of the in-plane N\'{e}el vector.) The Berry curvature of band $n = 1$ is dominantly positive, and the Chern number $C_1$ jumps from $0$ to $-1$. Thus, the magnetic phase transition is accompanied by a topological phase transition and $|\kappa_{xy}|$ is abruptly increased.

Ramping up the magnetic field further, the large Berry curvature around $\Gamma$ [cf.~Fig.~\ref{fig:spectra_berryc}(b)] becomes redistributed to high-energy magnons [cf.~Fig.~\ref{fig:spectra_berryc}(c)], with the consequence that $|\kappa_{xy}|$ decreases with the $B$ field [cf.~Fig.~\ref{fig:the}(b)].

The second topological phase transition is attributed to a band inversion. More precisely, at $g \mub B^\text{(t)}_2 = \SI{1.901}{\milli\electronvolt}$ the two bands intersect again and their Chern numbers are interchanged, that is $C_1 = -1 \to C_1 = +1$. This band inversion occurs near the BZ edge: just before $B^\text{(t)}_2$, e.\,g., $|g \mub B_z| = \SI{1.8}{\milli\electronvolt}$ the dominating positive Berry curvature appears near the BZ edge and is spread along $k_x$ [red in Fig.~\ref{fig:spectra_berryc}(c)]. And after the transition, e.\,g., at $|g \mub B_z| = \SI{2}{\milli\electronvolt}$, this dominating $\varOmega_{n\vect{k}}$ has changed sign [blue in Fig.~\ref{fig:spectra_berryc}(d)]. As a consequence, the band inversion manifests itself in $\kappa_{xy}$ prominently at elevated temperatures, for which it even causes sign changes [cf.~red line in Fig.~\ref{fig:the}(b)]. 

The band inversion is reversed again ($C_1 = +1 \to C_1 = -1$) at $g \mub B^\text{(t)}_3 = \SI{13.368}{\milli\electronvolt}$, again most clearly seen in $\kappa_{xy}$ at \SI{30}{\kelvin}, which, as before, features a sign change.
Approaching $B^\text{(t)}_3$ the elongated distribution of the Berry curvature seen for $B^\text{(t)}_2$ becomes concentrated around the K and K' points, and the band inversion then occurs at these points at the BZ edge (not shown). In short, the higher the temperature (but still well below the ordering temperature), the stronger $\kappa_{xy}$ reflects the topological phase transitions.

The second-order magnetic SF-FP phase transition at $g \mub B^\text{(m)}_2 = \SI{23.107}{\milli\electronvolt}$, also identified by a jump in the susceptibility, shows clear temperature-dependent signatures in $\kappa_{xy}$ [Fig.~\ref{fig:the}(b)]. On the one hand, the dominating positive contribution of the Berry curvature is located at the BZ edges (magnons with higher energies), on the other hand a small annular, negative contribution shows up near the BZ center (low-energy magnons) [Fig.~\ref{fig:spectra_berryc}(e)]. Thus, the weighting between these competing contributions can be altered by the occupation of the respective magnon states and, therefore, by the temperature. To be more specific, low temperatures freeze out the high-energy contribution, allowing the small low-energy contribution to dominate in the transport and leading to a peak with a sign change in $\kappa_{xy}$. At elevated temperatures, however, magnons with positive, eventually dominating $\varOmega_{n\vect{k}}$ are significantly populated. Since the high-energy contribution, being induced by the topological phase transition, exists independently of the magnetic phase transition, it does not show up as a pronounced peak.

At the transition point the in-plane Néel vector vanishes and the out-of-plane spin components have reached their maxima [inset in Fig.~\ref{fig:the}(a)]. The FP phase is hence characterized by a saturated (classical) magnetization [cf.~Fig.~\ref{fig:the}(a)]. Beyond this second-order transition the magnetic field shifts both bands to higher energies, thereby suppressing thermal transport ($\kappa_{xy} \to 0$) [Fig.~\ref{fig:the}(b)].

Based on the above we conclude that $\kappa_{xy}$ exhibits clear signatures of magnetic phase transition at low temperatures and of topological phase transitions at higher temperatures.

\paragraph{Thermal Hall magnetoconductivity.}
\begin{figure}
    \centering
    \includegraphics[width=\linewidth]{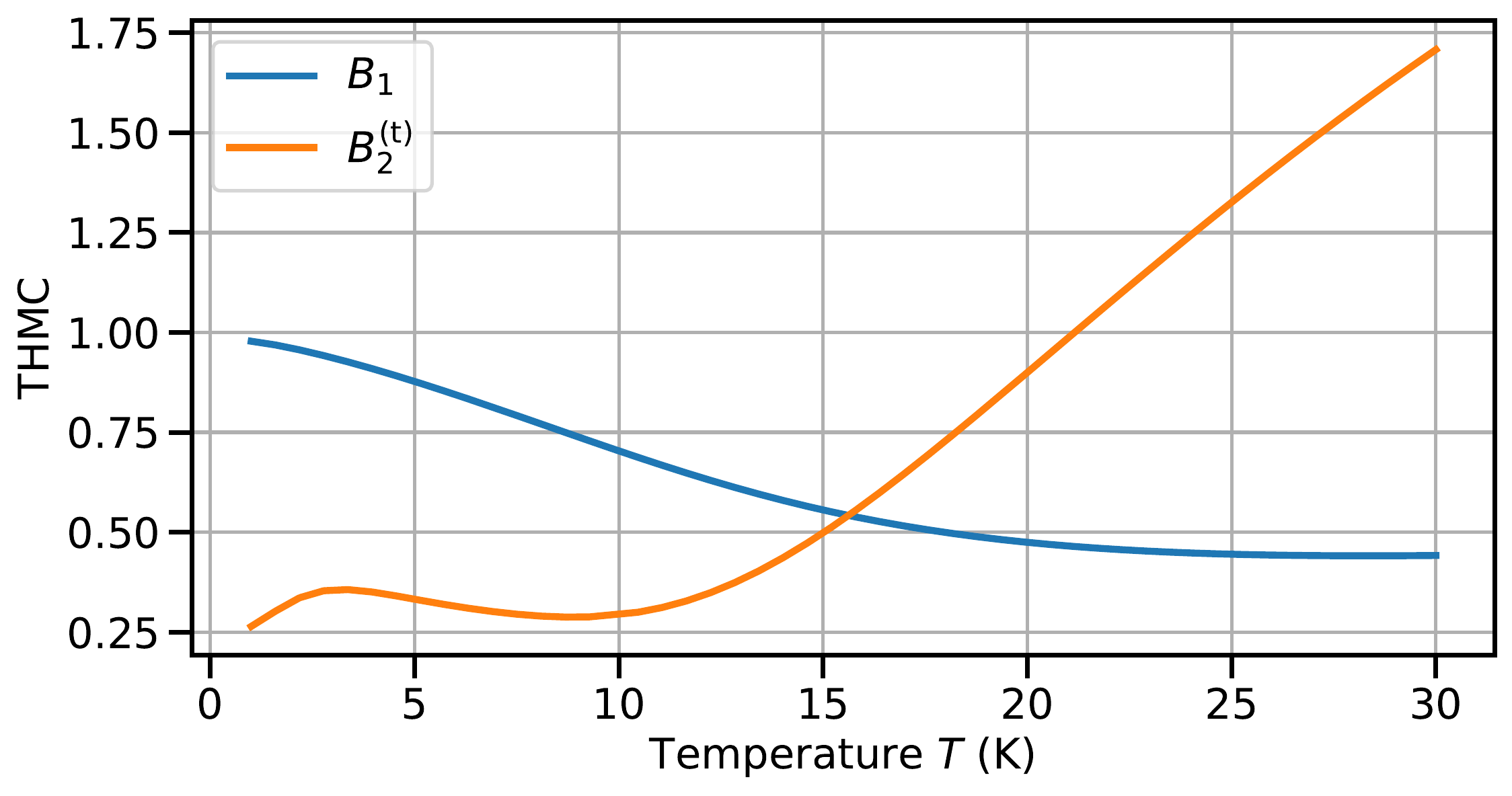}
    \caption{Thermal Hall magnetoconductivity THMC as a function of temperature $T$ at the AFM-SF transition (blue line) and the second topological phase transition (orange line).}
    \label{fig:thmc}
\end{figure}
The previous analysis revealed the need for a quantity that precisely measures the sensitivity of $\kappa_{xy}(B_z)$ on the phase transitions.
In analogy to the magnetoresistance, we define the thermal Hall magnetoconductivity (THMC)
\begin{align}
    \mathrm{THMC} = \abs{\frac{
        \kappa_{xy}(\bar{B_z} + \varDelta B_z) - \kappa_{xy}(\bar{B_z} - \varDelta B_z)
    }{
        \kappa_{xy}(\bar{B_z} + \varDelta B_z) + \kappa_{xy}(\bar{B_z} - \varDelta B_z)
    }}.
\end{align}
By definition the THMC corresponds to the relative change of $\kappa_{xy}$ upon the phase transition at $\bar{B}_z$.
In Fig.~\ref{fig:thmc} the THMC is shown with temperature for (i) the AFM-SF transition (blue line) and (ii) for the topological phase transition at $B^\text{(t)}_2$ (orange line) \footnote{We have set $\bar{B_z} = B_1$ and $\varDelta B_z = \SI{2}{\micro\electronvolt}$ for (i) and $\bar{B_z} = B_2^\text{(t)}$ and $\varDelta B_z = \SI{0.4}{\milli\electronvolt}$ for (ii)}.
For (i) the THMC is close to 1 near \SI{1}{\kelvin} and monotonically decreases with temperature.
(ii)~The topological phase transition shows the expected behavior, i.\,e., the THMC is small at low temperatures indicating that $\kappa_{xy}$ does not change by much, when the topological transition is crossed, but it escalates and takes values close to \SI{175}{\percent} at \SI{30}{\kelvin}.
Based on these results, the drastic changes of $\kappa_{xy}$ at the phase transitions can be exploited for a ``thermal Hall switch,'' in which the transverse heat current (or the transverse temperature gradient) is controlled by the external field.

\paragraph{Results for MnPS$_3$ on a substrate.}
A substrate or a heterostructure breaks the sublattice symmetry, which is mimicked by setting $A = \SI{0.1}{\milli\electronvolt}$.
There are three key differences to bulk MnPS$_3$ ($A = 0$): (i) the AFM-SF transition becomes continuous. While the A spins are only slightly deflected and rotated back to their $z$ orientation, the B spins perform a \SI{180}{\degree} rotation from being parallel to $-z$ via an in-plane geometry to eventually being parallel to $z$.
(ii) The nonmagnetic atoms, which are responsible for $A \neq 0$, break an effective time-reversal symmetry $PT$ and a thermal Hall effect in a collinear antiferromagnet without a magnetic field ensues.
A similar situation has been reported for the anomalous Hall effect in an electronic system \cite{Smejkal2020}.
$\kappa_{xy}(B_z = 0)$ can become large in materials with sizable $J_a$ that is responsible for strong nonzero Berry curvature associated with the low-energy magnon states near $\Gamma$ (cf. SM \cite{Supplement}).
(iii) $A$ opens a trivial gap in the FM phase and it dominates over $J_a$. Since the AF phase is always trivial, there are no topological phase transitions.
We present the magnon spectra, Berry curvature, thermal Hall effect, and heat capacity for $A = \SI{0.1}{\milli\electronvolt}$ in the SM \cite{Supplement}.
The critical fields are arranged in Table~\ref{tab:critfields}.

\paragraph{Wrap up.}
Our theoretical investigation of the temperature and magnetic-field dependence of the transversal heat conductivity $\kappa_{xy}$ of a honeycomb magnet proves that $\kappa_{xy}$ is very sensitive to the magnetic structure at low temperatures: it exhibits pronounced peaks at the magnetic phase transitions, but is rather unaffected by topological phase transitions. Conversely, $\kappa_{xy}$ traces the topological phase transitions at high temperatures, but is insensitive to the magnetic transitions. Its reading may change several orders near a phase transition and it may also change sign. To paraphrase, magnetic and topological phase transition cause distinct signatures in $\kappa_{xy}$, the measurement of which may be used to identify the phase transitions.
On the other hand, the strong change under the phase transitions may be exploited as a ``thermal Hall switch'' in which the transport properties are manipulated by external means.

Detecting topological (edge) magnons is more difficult than for electrons, since transport of bosons is not quantized --- what is a clear signature of nontrivial topology in electronic systems. Instead, $\kappa_{xy}(B)$ may be investigated as a `substitute,' its prominent features evidencing the existence of topological magnons.
Although there are other sources of drastic changes in $\kappa_{xy}(B)$, a combination with measurements of, e.\,g., heat capacity $C_V(B)$, which is insensitive to topology, could be used to verify the topological nature of the signatures (proof of concept in SM \cite{Supplement}).

Our findings call for experimental validation. The numerical results for MnPS$_3$ suggest that $\kappa_{xy}$ lies within the experimentally accessible range. We point out that extraordinarily high fields would be required for mapping the entire phase diagram. Nonetheless, the antiferromagnet--spin-flop transition and the topological transition at \SI{2.202}{\milli\electronvolt} are experimentally amenable.

\begin{acknowledgments}
\paragraph{Acknowledgments.}
This work is supported by CRC/TRR 227 of Deutsche Forschungsgemeinschaft (DFG).
\end{acknowledgments} 


\bibliography{short,newrefs}

\end{document}